\documentclass[prd,aps,onecolumn,showpacs,preprintnumbers,amsmath,amssymb]{revtex4}
\usepackage{amsmath}
\usepackage{graphicx}
\usepackage{dcolumn}
\usepackage{bm}
\usepackage{amssymb}
\usepackage{latexsym}

\def\be{\begin{equation}}
\def\ee{\end{equation}}
\def\bea{\begin{eqnarray}}
\def\eea{\end{eqnarray}}
\newcommand{\hphi}{\hat{\phi}}
\newcommand{\tphi}{\tilde{\phi}}
\newcommand{\tchi}{\tilde{\chi}}

\begin{document}
\markright{CYCU-HEP-10-07}

\title{Perturbations in Matter Bounce with Non-minimal Coupling}

\author{Taotao Qiu\footnote{qiutt@mail.ihep.ac.cn} and Kwei-Chou Yang}

\affiliation{Department of Physics, Chung-Yuan Christian University,
Chung-li, Taiwan 320}

\begin{abstract}

In this paper, we investigate the perturbations in matter bounce
induced from Lee-Wick lagrangian with the involvement of non-minimal
coupling to the Einstein Gravity. We find that this extra
non-minimal coupling term can cause a red-tilt on the primordial
metric perturbation at extremely large scales. It can also lead to
large enhancement of reheating of the normal field particles
compared to the usual minimal coupling models.
\end{abstract}

\maketitle

\section{introduction}
Non-singular bouncing cosmology \cite{Mukhanov:1991zn} has become
one of the important alternative scenarios of inflation
\cite{Guth:1980zm} as a description of the early universe. In terms
of a bounce scenario, the universe travels from a contracting phase
to an expanding phase through a non-vanishing minimal size, avoiding
the singularity problem which plagues the Standard Big Bang theory
\cite{Hawking:1973uf} or inflationary theory \cite{Borde:1993xh}.
Moreover, a bounce can occur much below the Planck scale,
eliminating the ``Transplanckian" problem of which the wavelength of
the fluctuation mode we see today will be even smaller than Planck
scale and thus in ``the zone of ignorance" where high energy effects
are robust and Einstein equations might be invalid
\cite{Martin:2000xs}. In bounce models, fluctuations were generated
in contracting phase and transferred into expanding phase through
bouncing point, which can give rise to a scale invariant power
spectrum as expected by the current observational data
\cite{Wands:1998yp}. In order to connect the perturbations in
contracting phase and expanding phase at the crossing point, the
joint condition of Hwang-Vishniac \cite{Hwang:1991an}
(Deruelle-Mukhanov \cite{Deruelle:1995kd}) can be applied. Besides,
bouncing scenario can also be originated from non-relativistic
gravity theories \cite{Cai:2009in} and give rise to gravitational
waves with running signature and large non-Gaussianities
\cite{Cai:2008ed}.

In the previous studies, a kind of ``matter bounce" has been
proposed where before and after the bounce, the universe could
behave as non-relativistic matter. This scenario can be realized by
a kind of ``Lee-Wick" (LW) Lagrangian which contains higher
derivatives of the field, or equivalently by two scalar fields with
an opposite sign of potential. This scenario can give rise to scale
invariant power spectrum at large scales, which is consistent with
the data \cite{Cai:2008qw}. However, one may expect that there are
some non-minimal couplings of the matter in the universe to Einstein
Gravity in the early universe \footnote{There are plenty of
non-minimal coupling theories in the literature, such as
scalar-tensor theories \cite{Fujii}, Brans-Dicke theory
\cite{Brans:1961sx} or dilaton theory \cite{Gasperini:1992em}. See
\cite{Nojiri:2005vv} for Gauss-Bonnet type of non-minimal couplings.
For comprehensive reviews see \cite{Sotiriou:2008rp}.}. In fact, it
is argued that the existence of the non-minimal coupling term is
required by the quantum corrections and renormalization
\cite{Chernikov:1968zm} in quantum gravity in curved space-time.
This effect may, by means of modifying Einstein equations, alter the
previous result and leave some signals on the power spectrum that
can be detected by observations such as CMB analysis. The
non-minimal coupling can also be reduced from higher dimensional
theories such as brane theory and can get rid of the Big Bang
Singularity, which lead to a bouncing universe \cite{Setare:2008qr},
or lead to the cosmic acceleration, which can be utilized as
inflaton in the early time \cite{Faraoni:1996rf} and dark energy at
current epoch \cite{Faraoni:2000wk}.

This paper aims at investigating the perturbations in the matter
bounce model involving the non-minimal coupling. The paper is
organized as follows: we first review the model of matter bounce in
Sec. II. After that, in Sec. III we take the non-minimal coupling
term into account. We investigate the perturbation through the
process in detail, and show the solutions for each stage. We also
analyze the particle production due to the resonance. All the
analytical calculations are done with help of numerical
computations. Finally Sec. IV contains conclusions and discussions.

\section{review of the matter bounce model}
In this section, we will start with the matter bounce model carried
out in \cite{Cai:2008qw}. This model consists of a single scalar
field with a high derivative term of a minus sign. Due to this term,
the equation of state (EoS) $w$ is possible to go down below $-1$
and violates the null energy condition, which behaves as a {\it
``Quintom"} matter \cite{Feng:2004ad} and makes it possible to
realize a bounce. It is useful to have such a term added in the
lagrangian. For example, in the well known Lee-Wick theory which was
constructed by T.D. Lee and G. C. Wick \cite{Lee:1969fy} (see also
\cite{Grinstein:2007mp} for the extensive ``Lee-Wick standard model"
), the higher derivative term is used to cancel the quadratic
divergence of the Higgs mass and thus address the ``hierarchy
problem". It is also possible to construct an ultraviolet complete
theory which preserves Lorentz invariance and unitarity in terms of
Lee-Wick theory \cite{vanTonder:2008ub}.

We begin with the Lagrangian in our model to be in the form:
\be\label{leewick} {\cal
L}=\frac{1}{2}\partial_{\mu}\hphi\partial^{\mu}\hphi-\frac{1}{2M^2}(\partial^2
\hphi)^2-\frac{1}{2}m^2\hphi^2-V(\hphi)~, \ee where $m$ is the mass
of the scalar field $\hphi$ and $V(\hphi)$ is the potential. A
higher derivative term with minus sign is introduced with some new
mass scale $M$. For the Higgs sector of Lee-Wick theory, the
hierarchy problem is solved if we require that $M\gg m$. After some
field rotations, one can write down the effective Lagrangian:
\be\label{leewick2} {\cal
L}=\frac{1}{2}\partial_{\mu}\phi\partial^{\mu}\phi-\frac{1}{2}
\partial_{\mu}\tphi\partial^{\mu}\tphi+\frac{1}{2}M^2\tphi^2-\frac{1}{2}m^2\phi^2-V(\phi,\tphi)~, \ee where $\tphi$ is
some auxiliary field and $\phi$ is defined as
$\phi\equiv\hphi+\tphi$  \footnote{People may worry about the
"ghost" behavior of the field $\tilde\phi$, which may cause some
instability problems \cite{Carroll:2003st}. That is true but it is
not the focus of our paper. Note that we are only interested in the
general picture of the bounce, and the auxiliary field is only used
as an explicit example for the bounce to happen. A more general
analysis will be taken into consideration in the future work. }.
Here the mass matrix of the fields has been diagonalized due to the
rotation. Usually there may be some interaction terms between the
two fields, or some higher order self-interaction terms like
$\phi^4$, $\tphi^4$, $\phi^2\tphi^2$ and so on, but here for
simplicity and without losing generality, we will have all of them
ignored by setting $V(\phi,\tphi)=0$. In framework of
Friedmann-Robertson-Walker (FRW) metric, \be ds^2=dt^2-a(t)^2d{\bf
x}^2~, \ee it is easy to get the Friedmann equation as: \be
\label{Heq}
H^2=\frac{\kappa^2}{3}\bigl[\frac{1}{2}{\dot\phi}^2-\frac{1}{2}{\dot\tphi}^2+\frac{1}{2}m^2\phi^2-\frac{1}{2}M^2\tphi^2\bigr]~,
\ee where $\kappa^2\equiv8\pi G$, and the equations of motion of the
two fields are \bea \label{KGeq}
\ddot\phi+3H\dot\phi+m^2\phi&=&0~, \nonumber\\
\ddot\tphi+3H\dot\tphi+M^2\tphi&=&0~, \eea respectively.

Let us now take a close look at how the model works to give rise to
a bouncing scenario. The bounce happens by the conditions $H=0$ and
$\dot H>0$, which requires that the total energy density vanish at
some point in the universe evolution. Starting off in the
contracting phase, both the two fields $\phi$ and $\tphi$ oscillate
around the extrema of their potential with a growing amplitude due
to the anti-friction of the negative Hubble parameter. Therefore,
the energy densities of both fields will grow as $a^{-3}(t)$, which
behaves like non-relativistic matter. Note that the energy density
of $\tphi$ is negative and its growth will cancel that of the total
energy density of the universe. At the beginning, we assume that it
is subdominant than that of $\phi$, but as $M\gg m$, it will grow
faster and eventually catch up, canceling all the positive energy
density with its kinetic energy overwhelming that of $\phi$, which
makes $\dot H>0$ and cause bounce to happen. Thus we can see that
before and after the bounce, the universe can be viewed as ``matter
domination" with the effective equation of state $w=0$, while in the
neighborhood of the bounce point, the equation of state goes down to
$-\infty$ \footnote{Remark 1: As is obvious, the divergent of EoS is
due to the vanishment of the total energy density of the universe
and is thus not avoidable. Fortunately,since EoS is not a physical
quantity, it is safe and will not cause any real instabilities.\\
Remark 2: Recently a paper \cite{Karouby:2010wt} pointed out that
this kind of model cannot be stable with the addition of regular
radiation in the contracting phase. Possible keys to this issue is
being under investigation.}. It is a good approximation to
parameterize $w$ in such a way for the following calculations.

In the following section, we will reinvestigate the scenario by
taking into account the coupling of the field to Einstein Gravity.
We assume that the normal scalar couples to the curvature through
some non-minimal coupling term such as $\frac{1}{2}\xi R\phi^2$. We
will analyze in detail the evolution of the perturbations in
presence of this term, including the metric perturbations as well as
the effects on particle production. To support our analytical
calculations, we will also perform the numerical investigations.

\section{adding non-minimal coupling term to the LW bounce}
\subsection{background}
In this section, we investigate the system with the inclusion of the
non-minimally coupled term. Generally, we can modify Lagrangian
(\ref{leewick}) in a very general form such as: \be {\cal
L}=\frac{1}{2}\partial_{\mu}\hphi\partial^{\mu}\hphi-\frac{1}{2M^2}(\partial^2
\hphi)^2-\frac{1}{2}m^2\hphi^2-V(\hphi)-f(R)h(\hphi)~, \ee where
$f(R)$ and $h(\hphi)$ are functions of the Ricci scalar $R$ and
scalar field $\hphi$, respectively. By writing it in terms of $\phi$
and $\tphi$ in analogy to (\ref{leewick2}), we can always have: \be
{\cal L}=\frac{1}{2}\partial_{\mu}\phi\partial^{\mu}\phi-\frac{1}{2}
\partial_{\mu}\tphi\partial^{\mu}\tphi+\frac{1}{2}M^2\tphi^2-\frac{1}{2}m^2\phi^2-V(\phi,\tphi)-f_1(R)h_1(\phi)-f_2(R)h_2(\tphi)~, \ee
where $f_1(R)$, $f_2(R)$, $h_1(\phi)$ and $h_2(\tphi)$ are inherited
from $f(R)h(\hphi)$ in the original Lagrangian. Usually, these
functions are very complicated and difficult to analyze. For the
sake of simplicity, we adopt a kind of parametrization form in which
the coupling of $\phi$ has a very simple form such as
$\frac{1}{2}\xi R\phi^2$ with a coefficient $\xi$, while that of
$\tphi$ vanishes. The parameterized Lagrangian is: \be\label{action}
{\cal S}=\int
d^4x\sqrt{-g}\bigl[\frac{R}{2\kappa^2}+\frac{1}{2}\partial_{\mu}\phi\partial^{\mu}\phi-\frac{1}{2}
\partial_{\mu}\tphi\partial^{\mu}\tphi+\frac{1}{2}M^2\tphi^2-\frac{1}{2}m^2\phi^2-\frac{1}{2}\xi R\phi^2\bigr]~.
\ee From this Lagrangian, we obtain the Einstein Equations as:
\be\label{modeinstein}
\frac{1-\kappa^2\xi\phi^2}{\kappa^2}G_{\mu\nu}=g_{\mu\nu}
[\frac{1}{2}{\dot\phi}^2-\frac{1}{2}{\dot\tphi}^2+\frac{1}{2}m^2\phi^2-\frac{1}{2}M^2\tphi^2\bigr]
-\nabla_\mu\phi\nabla_\nu\phi+\nabla_\mu\tphi\nabla_\nu\tphi-2\xi
g_{\mu\nu}({\dot\phi}^2+\phi\Box\phi)+2\xi(\nabla_\mu\phi\nabla_\nu\phi+\phi\nabla_\mu\nabla_\nu\phi)~,\ee
and the equations of motion of $\phi$ and $\tphi$: \bea
\ddot\phi+3H\dot\phi+m^2\phi+\xi R\phi&=&0~, \nonumber\\
\ddot\tphi+3H\dot\tphi+M^2\tphi&=&0~, \eea respectively. In deriving
these equations we have made use of the definition $R\equiv
g^{\mu\nu}R_{\mu\nu}=6(\dot H+2H^2)$.

Eq. (\ref{modeinstein}) yields equation for Hubble parameter
straightforwardly, which is: \be
H^2=\frac{\kappa^2}{3(1-\kappa^2\xi\phi^2)}
[\frac{1}{2}{\dot\phi}^2-\frac{1}{2}{\dot\tphi}^2+\frac{1}{2}m^2\phi^2-\frac{1}{2}M^2\tphi^2+6\xi
H\phi\dot\phi\bigr]~,\ee and the Ricci scalar is: \be
R=\frac{\kappa^2}{1-\kappa^2\xi\phi^2(1-6\xi)}\bigl[-(1-6\xi){\dot\phi}^2+2(1-3\xi)m^2\phi^2+{\dot\tphi}^2-2M^2\tphi^2\bigr]~.\ee

Generally, the additional term will affect the background evolution
by modifying the field equations. This may make the analysis
sophisticated and undoable. Moreover, if the modification is too
large, the bounce may even be prevented. However, for the case of
weak coupling, it is still suitable for us to preserve the bounce
and take assumption that the background evolution is hardly
affected. Thus it will be safe to use all the results obtained in
the above section without losing fidelity.

\begin{figure}[htbp]
\includegraphics[scale=0.4]{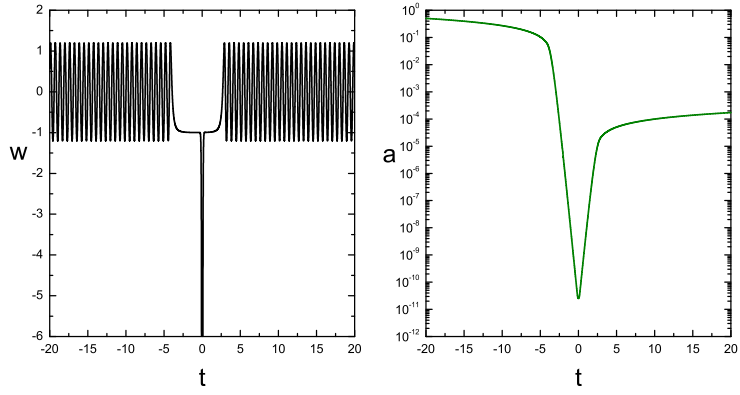}
\caption{(Color online) Evolution of the background equation of
state parameter $w$ and scale factor $a(t)$ in a non-minimal
coupling matter bounce model as a function of cosmic time $t$
(horizontal axis). The background fields are plotted in
dimensionless units by normalizing by the mass $M_{rec}=10^{-6}
m_{pl}$, and the time axis is displayed in units of $M_{rec}^{-1}$.
The mass parameters $m$ and $M$ are chosen to be $m=5.0 M_{rec}$ and
$M=10.0 M_{rec}$. The initial conditions are $\phi_i=-370.514
M_{rec}$, $\dot\phi_i=-6752.791 M_{rec}^2$,
$\tilde\phi_i=-4.818\times10^{-11} M_{rec}$,
$\dot{\tilde\phi}_i=-8.837\times10^{-11} M_{rec}^2$.} \label{eos}
\end{figure}

To be more accurate, we test the background evolution numerically.
We choose the parameters and initial conditions to be the same as in
\cite{Cai:2008qw}. We find that up to $|\xi|\sim 0.05$, the
background evolution will not be affected very much and a bounce can
still happen. But there may be a little difference, where the
universe may enter into a small stage with $w\simeq-1$ before
bounce, as have been shown in Fig. \ref{eos}. This is easy to
understand from the field equation. In the additional term, $\xi R$
can be viewed as an effective correction to the original mass
squared $m^2$. At the initial stage where $R$ is not very large, the
correction is negligible, and the field evolves as it does without
the additional term. But as the universe evolves like pressureless
matter, $R$ will get larger and larger. When the Hubble parameter is
about to its maximal value, $H\sim m$, the second scalar $\tphi$
will begin to dominate the universe where $\dot H$ transfers from
negative to positive. At this moment, from the definition of $R$, we
can see that it suddenly jumps by the amount of $2|\dot H|$. This
behavior makes the energy density dominate the universe again for a
short while, leading to $w\simeq-1$ in this period. As the energy
density of $\tphi$ grows up and totally dominates the universe, the
equation of state goes down to be below $-1$ and drives the bounce
to happen. At the bouncing point where the total energy density of
the universe vanishes, one can see from Fig. \ref{eos} that the
equation of state goes down to negative infinity, and as stated
before (in the footnote), it is not a real divergence. After the
bounce, the EoS will come up to above $-1$ again as the time
reversal of the pre-bounce process.

\subsection{perturbation}
In this section we will focus on the perturbation evolution of our
model. We begin with the perturbed metric, \be
ds^2=a(\eta)^2[(1+2\Phi)d\eta^2-(1-2\Psi)d{\bf x}^2]~,\ee in the
conformal form. Here $\eta\equiv\int\frac{dt}{a}$ is the conformal
time, and $\Phi$ and $\Psi$ are scalar perturbations of the metric.
Note that with existence of the non-minimal coupling term, $\Phi$
and $\Psi$ are no longer equivalent. Similarly, the fields can be
represented as homogeneous and fluctuation parts as: \bea
\phi&=&\phi_0+\delta\phi~,\nonumber\\
\tphi&=&\tphi_0+\delta\tphi~.\eea In the following, we will neglect
the subscript ``0" and take $\phi$ and $\tphi$ as background
components. Then we will solve the perturbed Einstein Equations to
see their evolutions.

\subsubsection{general solution}

From the perturbed Einstein Equation, \be \delta
G_{\mu\nu}=-\kappa^2\delta T_{\mu\nu}~,\ee we obtain the following
equations for perturbation variables: \be\label{offdiagonal}
\Phi=\Psi-\frac{\delta F}{F}~,\ee \be\label{0i}
\dot\Psi+(H-\frac{\dot
F}{2F})\Phi=\frac{\kappa^2}{2F}(-\dot\phi\delta\phi+\dot\tphi\delta\tphi)+\frac{\dot{\delta
F}}{2F}+\frac{H\delta F}{2F}~,\ee \bea & &\ddot\Psi+3(H+\frac{\dot
F}{2F})\dot\Psi+\left\{\frac{\kappa^2}{F}\bigl[(\dot\phi^2+12\xi
H\phi\dot\phi)+2\xi\dot\phi^2\bigr]-(2\dot
H+3H^2)-\frac{\kappa^2m^2\phi^2}{2F}-\frac{\ddot
F}{F}\right\}\Psi\nonumber\\ &=&\frac{\delta
F}{F}\left\{\frac{3}{2}(2\dot
H+3H^2)-\frac{\kappa^2}{F}\bigl[(\frac{\dot\phi^2}{2}+6\xi
H\phi\dot\phi)+2\xi\dot\phi^2\bigr]+\frac{k^2}{2a^2}+\frac{\dot
F^2}{2F^2}\right\}+\frac{\ddot{\delta F}}{2F}+\frac{\dot
F\dot{\delta F}}{2F^2}\nonumber\\
&+&(\frac{\kappa^2m^2\phi}{F}+\frac{3\kappa^2H\dot\phi}{F})\delta\phi-\frac{\kappa^2}{F}(\frac{\dot\phi}{2}+2H\xi\phi)\dot{\delta\phi}~,\eea
\bea &
&\ddot{\delta\phi}+3H\dot{\delta\phi}+\left\{\bigl(\frac{k}{a}\bigr)^2+m^2+6\xi(\dot H+2H^2)\right\}\delta\phi\nonumber\\
&=&6\xi\phi\ddot\Psi+24\xi H\phi\dot\Psi+3\dot\phi\dot\Psi+6\xi
H\phi\dot\Phi+\dot\phi\dot\Phi+\frac{4\xi
k^2\phi\Psi}{a^2}+2\xi\phi\bigl[6(\dot
H+2H^2)-\bigl(\frac{k}{a}\bigr)^2\bigr]\Phi-2(m^2\phi+\xi
R\phi)\Phi~,\eea where we have defined $F\equiv1-\kappa^2\xi\phi^2$.

Since the equations are too sophisticated and there is no hope to
solve it directly, we resort to the Einstein frame by a conformal
transformation of the metric \cite{Mukhanov:1990me}: \be
\hat{g}_{\mu\nu}=Fg_{\mu\nu}~.\ee The action (\ref{action}) is then
transformed into the form: \be {\cal S}=\int
d^4x\sqrt{-\hat{g}}\bigl[\frac{\hat{R}}{2\kappa^2}+\frac{1-(1-6\xi)
\kappa^2\xi\phi^2}{2F^2}\hat{\partial}_{\mu}\phi\hat{\partial}^{\mu}\phi
-\frac{1}{2F}\hat{\partial}_{\mu}\tphi\hat{\partial}^{\mu}\tphi+
\frac{1}{F^2}(\frac{1}{2}M^2\tphi^2-\frac{1}{2}m^2\phi^2)\bigr]~.
\ee

We introduce two new variables, $\chi$ and $\tchi$, which are
defined as: \bea \chi&\equiv&\int\frac{\sqrt{1-(1-6\xi)
\kappa^2\xi\phi^2}}{F}d\phi~,\nonumber\\
\tchi&\equiv&\int\frac{1}{F}d\tphi~,\eea and rewrite the above
action as: \be {\cal S}=\int
d^4x\sqrt{-\hat{g}}\bigl[\frac{\hat{R}}{2\kappa^2}+\frac{1}{2}\hat{\partial}_{\mu}\chi\hat{\partial}^{\mu}\chi
-\frac{1}{2}\hat{\partial}_{\mu}\tchi\hat{\partial}^{\mu}\tchi+
\frac{1}{F^2}(\frac{1}{2}M^2\tphi(\tchi)^2-\frac{1}{2}m^2\phi(\chi)^2)\bigr]~.
\ee The perturbed metric will be:  \be
ds^2=\hat{a}^2(\eta)[(1+2\hat{\Phi})d\eta^2-(1-2\hat{\Psi})d{\bf
x}^2]~,\ee where $\hat{a}(\eta)=\sqrt{F}a(\eta)$.

For the perturbed metric in Einstein Frame where the off-diagonal
components of the perturbed Einstein equations vanishes, we have
$\hat{\Phi}=\hat{\Psi}$. This can also be obtained when we consider
up to the first order approximation of the perturbed metric, which
gives $\hat{\Phi}=\Phi+\frac{\delta F}{2F}$, and
$\hat{\Psi}=\Psi-\frac{\delta F}{2F}$ where relation between $\Phi$
and $\Psi$ has already been given in Eq. (\ref{offdiagonal}).
Moreover, Eq. (\ref{0i}) gives the direct relationship between
$\hat{\Phi}$ and the field perturbation, \be\label{relation}
\phi'\delta\phi-\tphi'\delta\tphi=\frac{2F^{\frac{3}{2}}(a\sqrt{F}\hat{\Phi})'}{\kappa^2a[1-(1-6\xi)
\kappa^2\xi\phi^2]}~,\ee which is convenient for us to calculate
$\hat{\Phi}$ in Einstein frame and then transform into the original
Jordan frame.

The equation for $\hat{\Phi}$ is: \bea\label{Phieq}
\hat{\Phi}''+2\bigl(\hat{\cal
H}-\frac{\chi''}{\chi'}\bigr)\hat{\Phi}'+2\bigl(\hat{\cal
H}'-\hat{\cal
H}\frac{\chi''}{\chi'}\bigr)\hat{\Phi}+k^2\hat{\Phi}=\kappa^2\bigl(2\hat{\cal
H}+\frac{\chi''}{\chi'}\bigr)\tchi'\delta\tchi~, \eea where
$\hat{\cal H}=\hat{a}'/\hat{a}$. The right-hand side (r.h.s.) of the
equation is assumed to be small and negligible except at the bounce
point. Moreover, it is also convenient to define the curvature
perturbation $\hat{\zeta}$ in terms of $\hat{\Phi}$: \be
\hat{\zeta}=\hat\Phi+\frac{\hat{\cal H}}{\hat{\cal H}^2-\hat{\cal
H}'}(\hat{\Phi}'+\hat{\cal H}\hat{\Phi})~,\ee which is expected to
be conserved on super-Hubble scales in the inflationary universe by
the condition that the entropy perturbations are small and
$(1+w)\neq0$: \be (1+w)\hat{\zeta}'=0~.\ee

At the next step we will solve Eq. (\ref{Phieq}) for each stage to
get the perturbations $\Phi$, $\Psi$ and $\delta\phi$, and see how
it is effected by the non-minimal coupling term.
\subsubsection{contracting phase}
\begin{figure}[htbp]
\includegraphics[scale=0.4]{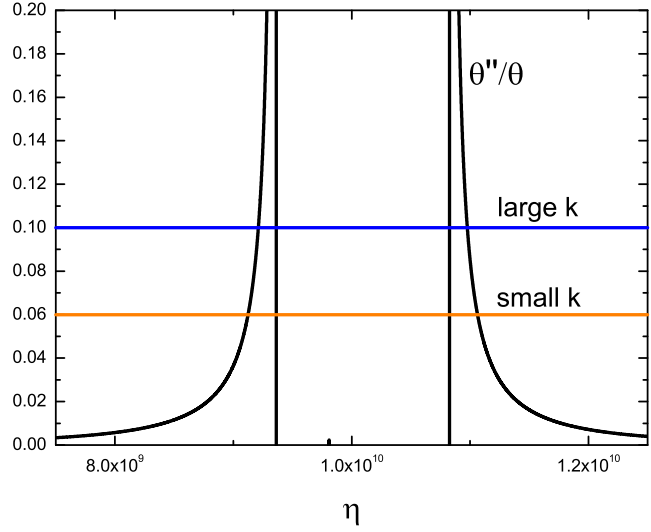}
\caption{(Color online) A sketch of the evolution of fluctuation
modes compared to the potential $\theta''/\theta$ in a bouncing
universe. The vertical axis is co-moving spatial coordinate, and the
horizontal axis is the conformal time. The orange line denotes the
fluctuations with small comoving wavenumber $k$ while the blue line
is for those with large $k$.}\label{potential}
\end{figure}

In contracting phase, the universe is dominated by the normal scalar
$\phi$. Since $F$ deos not vary significantly, we roughly have
$\tchi\sim\tphi$ which does not play an important role in the
evolution. Then by neglecting the r.h.s. of Eq. (\ref{Phieq}), we
consider it as a homogeneous equation. It is useful to define a new
variable $u\equiv\hat{a}\hat{\Phi}/\chi'$ and rewrite Eq.
(\ref{Phieq}) as: \be u''+(k^2-\frac{\theta''}{\theta})u=0~,\ee
where $\theta=\hat{\cal H}/(\hat{a}\chi')$. The solution can be
split into two limits: short-wavelength $(k^2\gg\theta''/\theta)$
and long-wavelength $(k^2\ll\theta''/\theta)$ where
$\theta''/\theta$ is the potential. For the short-wavelength
perturbations where the potential term can be neglected, we have:
\be\label{short} u=Ce^{ik\eta}+c.c.~, \ee where $C$ is a constant
and $c.c.$ stands for the complex conjugate. For long-wavelength
perturbations, one gets: \be\label{long}
u=\theta(S+D\int\frac{d\eta}{\theta^2})~.\ee From the above argument
of $\Phi$ and $\Psi$ as well as the definition of $u$, one can find
the solutions for perturbation variables in terms of $u$ which are:
\bea\label{phi} \Phi&=&\frac{u\hat{\cal
H}}{\hat{a}^2\theta}-\frac{F'}{2\hat{a}F(\hat{\cal H}^2-\hat{\cal
H}')}(\frac{u\hat{\cal
H}}{\hat{a}\theta})'~,\\
\label{psi} \Psi&=&\frac{u\hat{\cal
H}}{\hat{a}^2\theta}+\frac{F'}{2\hat{a}F(\hat{\cal H}^2-\hat{\cal
H}')}(\frac{u\hat{\cal
H}}{\hat{a}\theta})'~, \\
\label{deltaphi}\delta\phi&=&\frac{\phi'}{\hat{a}(\hat{\cal
H}^2-\hat{\cal H}')}(\frac{u\hat{\cal H}}{\hat{a}\theta})'~,\eea
where in the last formula we have neglected the contribution arising
from $\delta\tphi$ in (\ref{relation}), and we have used the
approximation $\chi'^2\simeq2(\hat{\cal H}^2-\hat{\cal
H}')/\kappa^2$. Substituting (\ref{short}) or (\ref{long}) to
(\ref{phi}), (\ref{psi}) and (\ref{deltaphi}), we get these
variables in Jordan frame: \bea
\Phi&=&C\sqrt{\frac{2}{\kappa^2}}\frac{e^{ik\eta}}{\sqrt{-\stackrel{\circ}{\hat{H}}}}
\left\{-\stackrel{\circ}{\hat{H}}-\frac{\stackrel{\circ}{F}}{2F}
\bigl(\hat{H}+\frac{\stackrel{\circ\circ}{\hat{H}}}{2\stackrel{\circ}{\hat{H}}}+\frac{ik}{\hat{a}}\bigr)\right\}+c.c.~,\\
\Psi&=&C\sqrt{\frac{2}{\kappa^2}}\frac{e^{ik\eta}}{\sqrt{-\stackrel{\circ}{\hat{H}}}}
\left\{-\stackrel{\circ}{\hat{H}}+\frac{\stackrel{\circ}{F}}{2F}
\bigl(\hat{H}+\frac{\stackrel{\circ\circ}{\hat{H}}}{2\stackrel{\circ}{\hat{H}}}+\frac{ik}{\hat{a}}\bigr)\right\}+c.c.~, \\
\delta\phi&=&C\stackrel{\circ}{\phi}\sqrt{\frac{2}{\kappa^2}}\frac{e^{ik\eta}}{\sqrt{-\stackrel{\circ}{\hat{H}}}}
\bigl(\hat{H}+\frac{\stackrel{\circ\circ}{\hat{H}}}{2\stackrel{\circ}{\hat{H}}}+\frac{ik}{\hat{a}}\bigr)+c.c.\eea
for the short-wavelength case, where $\hat{H}$ is the Hubble
parameter in Einstein Frame and the symbol $``\circ''$ denotes
cosmic time derivative in Einstein frame (for details see Appendix
A), or \bea
\Phi&=&(\frac{H}{aF}+\frac{\dot F}{aF^2})S+\frac{2}{\kappa^2}D\bigl[\frac{1}{F}(\frac{1}{a}\int aFdt)\dot{}-\frac{\dot F}{aF^2}\int aFdt\bigr]~,\\
\Psi&=&\frac{H}{aF}S+\frac{2}{\kappa^2}D\bigl[\frac{1}{F}(\frac{1}{a}\int aFdt)\dot{}\bigr]~,\\
\delta\phi&=&-\frac{\dot\phi}{aF}(S-\frac{2}{\kappa^2}D\int
aFdt)\eea for the long-wavelength case. The sketch plot of
fluctuation modes compared to $\theta''/\theta$ are shown in Fig.
\ref{potential}.

For the case that the universe evolves with some constant equation
of state $w$, the scale factor can be parameterized as: \be a(t)\sim
t^p~,\ee where $p$ is defined as $p\equiv2/[3(1+w)]$. This will be
the case in the whole process except for the bouncing point. During
the contracting phase before the non-minimal term dominates the
universe, the fields oscillate around their extrema which made the
universe behave like non-relativistic matter. The average value of
the equation of state of the universe is $<w>=0$ (see from Fig.
\ref{eos}), and the scale factor $a(t)$ scales as $t^{2/3}$. Since
in this period the factor $F$ does not evolve significantly, it can
be approximately viewed as a constant. Therefore the scaling of
$\Phi$, $\Psi$ and $\delta\phi$ are given as:
\label{solution1}\bea \Phi&\simeq&\Psi\simeq C\sqrt{\frac{4}{3\kappa^2}}\frac{1}{t}e^{-i\frac{k}{a_0}t^\frac{1}{3}}+c.c.~, \\
\delta\phi&\simeq&\frac{2C}{\kappa^2\sqrt{F}}(-\frac{1}{3t}+\frac{ik}{a_0t^\frac{2}{3}})e^{-i\frac{k}{a_0}t^\frac{1}{3}}+c.c.~\eea
for short-wavelength case, and
\label{solution2}\bea \Phi&\simeq&\Psi\simeq\frac{2S_c^-}{3a_0Ft^\frac{5}{3}}+\frac{6D_c^-}{5\kappa^2}~, \\
\delta\phi&\simeq&-\frac{S_c^-}{a_0F}\sqrt{\frac{4}{3\kappa^2}}t^{-\frac{5}{3}}+\frac{6D_c^-}{5\kappa^2}\sqrt{\frac{4}{3\kappa^2}}~\eea
for long-wavelength case, where $a_0$ is some normalization
constant. Here the subscript $``c"$ denotes the contracting phase
and superscript $``-"$ means the matter-like region which happens
earlier. Note that $\Phi$, $\Psi$ and $\delta\phi$ have magnitudes
of the same order which should be much less than $m_{pl}$, so the
difference between $\Phi$ and $\Psi$ (proportional to $\delta\phi$)
will be severely suppressed.

However, when the non-minimal term begins to dominate the universe,
the equation of state approaches to $-1$. During this period one
could define a slow roll parameter $\epsilon\equiv-\dot H/H^2$ which
should be very small. Thus we can have $a=a_1t^{1/\epsilon}$ and
$H=(\epsilon t)^{-1}$. Moreover, we have from the Einstein Equations
that $\dot\phi\sim\sqrt{\frac{4}{3\kappa^2}}\frac{1}{(t-t_{\ast})}$
and $\phi\sim\sqrt{\frac{4}{3\kappa^2}}\ln{(t-t_{\ast})}$, which
results in $F=1-\frac{4\xi\phi_0}{3}(\ln{(t-t_{\ast})})^2$, where
$\phi_0$ is some constant. From this we can have:
\label{solution3}\bea \Phi&\simeq&\Bigl(\frac{1}{a_1\epsilon
(t-t_{\ast})^{1+\frac{1}{\epsilon}}[1-\frac{4\xi\phi_0}{3}(\ln{(t-t_{\ast})})^2]}-\frac{8\xi\phi_0\ln{(t-t_{\ast})}}{3a_1(t-t_{\ast})^{1+\frac{1}{\epsilon}}[1-\frac{4\xi\phi_0}{3}(\ln{(t-t_{\ast})})^2]^2}\Bigr)S_c^+\nonumber\\
&+&\frac{2}{\kappa^2}\{\epsilon\Bigl(\frac{1}{1+\epsilon}+\frac{8\xi\phi_0[\epsilon-(1+\epsilon)\ln{(t-t_{\ast})}]}{3(1+\epsilon)^3[1-\frac{4\xi\phi_0}{3}(\ln{(t-t_{\ast})})^2]}\Bigr)\nonumber\\
&+&\frac{8\epsilon\xi\phi_0\ln{(t-t_{\ast})}}{3[1-\frac{4\xi\phi_0}{3}(\ln{t})^2]}\Bigl(\frac{1}{1+\epsilon}-\frac{8\epsilon\xi\phi_0[\epsilon+(1+\epsilon)\ln{(t-t_{\ast})}]}{3(1+\epsilon)^3[1-\frac{4\xi\phi_0}{3}(\ln{(t-t_{\ast})})^2]}\Bigr)\}D_c^+~,\\
\Psi&\simeq&\frac{S_c^+}{a_1\epsilon
(t-t_{\ast})^{1+\frac{1}{\epsilon}}[1-\frac{4\xi\phi_0}{3}(\ln{(t-t_{\ast})})^2]}+\frac{2\epsilon}{\kappa^2}\Bigl(\frac{1}{1+\epsilon}+\frac{8\xi\phi_0[\epsilon-(1+\epsilon)\ln{(t-t_{\ast})}]}{3(1+\epsilon)^3[1-\frac{4\xi\phi_0}{3}(\ln{(t-t_{\ast})})^2]}\Bigr)D_c^+~,\\
\delta\phi&\simeq&-\frac{\sqrt{\frac{4}{3\kappa^2}}}{a_1(t-t_{\ast})^{1+\frac{1}{\epsilon}}[1-\frac{4\xi\phi_0}{3}(\ln{(t-t_{\ast})})^2]}\nonumber\\
&
&\left\{S_c^+-\frac{2a_1\epsilon(t-t_{\ast})^{1+\frac{1}{\epsilon}}}{\kappa^2(1+\epsilon)^3}\Bigl((1+\epsilon)^2[1-\frac{4\xi\phi_0}{3}(\ln{(t-t_{\ast})})^2]-\frac{8\epsilon\xi\phi_0}{3}[\epsilon+(1+\epsilon)\ln{(t-t_{\ast})}]\Bigr)D_c^+\right\}~\eea
for long-wavelength case, where the superscript $``+"$ stands for
the deflationary region which comes later.
\subsubsection{bouncing phase}

When the universe is dominated by the auxiliary scalar $\tphi$, the
equation of state drops down to below $-1$, and the total energy
density of the unverse turns to decrease. When it goes to zero a
bounce will happen. It is rather complicated to solve the equation
(\ref{Phieq}) directly, however, we can make some modeling of the
evolution to have it simplified without losing fidelity.
Generalizing the parametrization method in \cite{Cai:2008qw} (see
also \cite{Cai:2007zv}), we parameterize the Hubble parameter near
the bouncing point of the form: \be H=\alpha_1 t+\beta_1
t^3+...~,\ee with positive constants $\alpha_1$ and $\beta_1$ of
proper dimensions whose magnitudes are determined by the
microphysics of the bounce. Note that it is unnecessary to contain
terms with even power-laws of $t$. In our case, it can be estimated
that $\alpha_1\sim m^2$. From the relations of variables between
Jordan and Einstein frames, we can also obtain the approximate value
of $\hat{H}$ with respect to $\hat{t}$ up to first order of $t$: \be
\hat{H}=\alpha\hat{t}+~ higher~order~terms~,\ee where
$\alpha\simeq4\alpha_1/9$. Since this parametrization is only valid
during bounce phase where we can neglect the higher order terms of
$|\eta-\eta_B|$ with $\eta_B$ being the comoving time at the bounce
point, we obtain the equation for $\hat{\Phi}$ of this phase as: \be
\hat{\Phi}''+2y(\eta-\eta_B)\hat{\Phi}'+(k^2+\frac{2y}{3})\hat{\Phi}\simeq0~,\ee
where $y\equiv12\alpha a_B^2/\pi$ and $a_B$ is the scale factor at
the bounce point. The solution is: \be
\hat{\Phi}=e^{-y(\eta-\eta_B)^2}\left\{EH_l(\sqrt{y}(\eta-\eta_B))+F_1F_1\bigl(-\frac{l}{2},\frac{1}{2},y(\eta-\eta_B)^2\bigr)\right\}~,\ee
where $H_l$ and ${}_1F_1$ denotes the $l$-th Hermite polynomial and
confluent hypergeometric function respectively, with \be
l\equiv-\frac{2}{3}+\frac{k^2}{2y}~.\ee The short- and
long-wavelength limits of the solution are quite different. For
short wavelength $(k^2\gg y)$, it reads: \be
\hat{\Phi}=e^{-\frac{y(\eta-\eta_B)^2}{2}}\{\frac{\hat{E}}{\sqrt{2l}}\sin{[k(\eta-\eta_B)]}+\hat{F}\cos{[k(\eta-\eta_B)]}\}~,\ee
while for long wavelength $(k^2\ll y)$, it is: \be
\hat{\Phi}=\hat{F}+\hat{E}\sqrt{y}(\eta-\eta_B)-(1+l)\hat{F}(\eta-\eta_B)^2+{\cal
O}((\eta-\eta_B)^3)~,\ee where
$\hat{E}\equiv-2^{1+l}\sqrt{\pi}E/\Gamma(-\frac{l}{2})$ and
$\hat{F}\equiv F+2^{l}\sqrt{\pi}E/\Gamma(\frac{1-l}{2})$.
\subsubsection{expanding phase}
The expanding phase can be viewed as time reversal process of the
contracting phase. In this period, both the two fields will roll
down along their potentials and begin to oscillate with a decaying
amplitude and redshifted energy density. Similar to the case of
contracting phase, at the beginning of expansion of the universe,
the non-minimal coupling term still remains large and dominate over
the mass term of the $\phi$ field, and thus $\phi$ can be
approximated as a slow-rolling field. This drives the total equation
of state $w$ to approach $-1$, and lead to a short period of
inflation. But soon, when the non-minimal coupling term becomes less
important, both fields will oscillate around their minimum, behaving
like non-relativistic matter again.

One can also get the short and long wavelength solutions of the
metric and field perturbations $\Phi$, $\Psi$ and $\delta\phi$ for
the expanding phase. We only care about the long-wavelength case
which make sense for observation today. As similar to the
contracting case just except for replacing the scripts, one gets:
\label{solution4}\bea \Phi&\simeq&\Bigl(\frac{1}{a_2\epsilon
(t-t_{\dag})^{1+\frac{1}{\epsilon}}[1-\frac{4\xi\phi_0}{3}(\ln{(t-t_{\dag})})^2]}-\frac{8\xi\phi_0\ln{(t-t_{\dag})}}{3a_2(t-t_{\dag})^{1+\frac{1}{\epsilon}}[1-\frac{4\xi\phi_0}{3}(\ln{(t-t_{\dag})})^2]^2}\Bigr)S_e^-\nonumber\\
&+&\frac{2}{\kappa^2}\{\epsilon\Bigl(\frac{1}{1+\epsilon}+\frac{8\xi\phi_0[\epsilon-(1+\epsilon)\ln{(t-t_{\dag})}]}{3(1+\epsilon)^3[1-\frac{4\xi\phi_0}{3}(\ln{(t-t_{\dag})})^2]}\Bigr)~\nonumber\\
&+&\frac{8\epsilon\xi\phi_0\ln{(t-t_{\dag})}}{3[1-\frac{4\xi\phi_0}{3}(\ln{(t-t_{\dag})})^2]}\Bigl(\frac{1}{1+\epsilon}-\frac{8\epsilon\xi\phi_0[\epsilon+(1+\epsilon)\ln{(t-t_{\dag})}]}{3(1+\epsilon)^3[1-\frac{4\xi\phi_0}{3}(\ln{(t-t_{\dag})})^2]}\Bigr)\}D_e^-~,\\
\Psi&\simeq&\frac{S_e^-}{a_2\epsilon
(t-t_{\dag})^{1+\frac{1}{\epsilon}}[1-\frac{4\xi\phi_0}{3}(\ln{(t-t_{\dag})})^2]}+\frac{2\epsilon}{\kappa^2}\Bigl(\frac{1}{1+\epsilon}+\frac{8\xi\phi_0[\epsilon-(1+\epsilon)\ln{(t-t_{\dag})}]}{3(1+\epsilon)^3[1-\frac{4\xi\phi_0}{3}(\ln{(t-t_{\dag})})^2]}\Bigr)D_e^-~,\\
\delta\phi&\simeq&-\frac{\sqrt{\frac{4}{3\kappa^2}}}{a_2(t-t_{\dag})^{1+\frac{1}{\epsilon}}[1-\frac{4\xi\phi_0}{3}(\ln{(t-t_{\dag})})^2]}\nonumber\\
&
&\left\{S_e^--\frac{2a_2\epsilon(t-t_{\dag})^{1+\frac{1}{\epsilon}}}{\kappa^2(1+\epsilon)^3}\Bigl((1+\epsilon)^2[1-\frac{4\xi\phi_0}{3}(\ln{(t-t_{\dag})})^2]-\frac{8\epsilon\xi\phi_0}{3}[\epsilon+(1+\epsilon)\ln{(t-t_{\dag})}]\Bigr)D_e^-\right\}~\eea
for the short inflationary period, and \label{solution5}\bea \Phi&\simeq&\Psi\simeq\frac{2S_e^+}{3a_3F(t-t_{\dag})^\frac{5}{3}}+\frac{6D_e^+}{5\kappa^2}~, \\
\delta\phi&\simeq&-\frac{S_e^+}{a_3F}\sqrt{\frac{4}{3\kappa^2}}(t-t_{\dag})^{-\frac{5}{3}}+\frac{6D_e^+}{5\kappa^2}\sqrt{\frac{4}{3\kappa^2}}~\eea
for the matter-like expanding phase. Here the subscript $``e"$
denotes expanding phase and superscripts $``-"$ and $``+"$ stand for
inflationary and matter-like regions separately.

\subsubsection{spectrum}
Having in hand the solutions in the above sections which stand for
different phases, now it is time to connect all of them using
matching conditions. According to \cite{Hwang:1991an} and
\cite{Deruelle:1995kd}, we can require that for each point which
joins two phases together, the three-metric as well as its extrinsic
curvature should be continuous. In conformal Newtonian gauge which
is used in this paper, this indicates that \be\label{matching}
\Psi\Big|_{\pm}=0~,~~~\frac{\nabla^2\Psi-3{\cal H}(\Psi'+{\cal
H}\Phi)}{3{\cal H}({\cal H}'-{\cal H}^2)}\Big|_{\pm}=0~,\ee where
$\pm$ means the difference before and after the transition point.
Substituting solutions for perturbations of each period into
(\ref{matching}) we can get the final results for them. From the
last paragraph we can see that at last the solution is divided by
two parts, one is constant ($D_e^+$ mode) and the other is decaying
($S_e^+$ mode). We are only interested in the first mode, which
dominates over the other one. Since the calculation is rather
straightforward and tedious, we only list the final result as
follows: \bea\label{final}
D_{e}^{+}&=&S_{c}^{-}({}^{(0)}G_{c}^{+}{}^{(0)}M_{e}^{+}+{}^{(0)}M_{c}^{+}{}^{(0)}N_{e}^{+})+D_{c}^{-}({}^{(0)}K_{c}^{+}{}^{(0)}M_{e}^{+}-{}^{(0)}N_{c}^{+}{}^{(0)}N_{e}^{+})\nonumber\\
&+&\frac{k^{2}}{a^{2}}[(S_{c}^{-}({}^{(1)}M_{c}^{+}{}^{(0)}N_{e}^{+}-{}^{(0)}M_{c}^{+}{}^{(1)}N_{e}^{+}-{}^{(1)}G_{c}^{+}{}^{(0)}M_{e}^{+}-{}^{(0)}G_{c}^{+}{}^{(1)}M_{e}^{+})\nonumber\\
&+&D_{c}^{-}({}^{(1)}N_{c}^{+}{}^{(0)}N_{e}^{+}-{}^{(1)}K_{c}^{+}{}^{(0)}M_{e}^{+}-{}^{(0)}K_{c}^{+}{}^{(1)}M_{e}^{+}+{}^{(0)}N_{c}^{+}{}^{(1)}N_{e}^{+})]\nonumber\\
&+&{\cal O}(\frac{k^4}{a^4})~.
\eea

For all the coefficients that appear above, we refer the readers to
Appendix B. These coefficients are all the specific value at the
joint point, so they are independent on $k$. Moreover, the mode in
contracting phase $D_c^-$ and $S_c^-$ can be easily read from the
initial conditions. If we set up with the Bunch-Davies vacuum which
implies $\Psi\sim k^{-\frac{3}{2}}t^{-1}$, we can obtain that
$D_c^-\sim k^{\frac{3}{2}}$ and $S_c^-\sim k^{-\frac{7}{2}}$.
Substituting them into (\ref{final}), we can see that at large
scales $D_c^-$ vanishes. The contributions of $S_c^-$ in the first
order in $k^2$ will be blue-tilted by exactly the right amount to
yield a scale invariant spectrum. This is the same behavior as the
normal case of matter bounce without non-minimal coupling.

However, different from the normal case, we find that the $S_c^-$
mode also has contributions in the zeroth order in $k^2$ to the
final spectrum. As we know, in the standard case of minimal
coupling, the constant mode after the transfer point cannot be
inherited from the running mode before the transfer. Actually, one
may find that the coefficients ${}^{(0)}M_{e}^{+}$ and
${}^{(0)}M_{c}^{+}$ vanish except for those terms containing $\xi$
(information from non-minimal coupling term), which is consistent
with its minimal coupling limit $\xi\rightarrow 0$, but when
non-minimal coupling terms are introduced, a mixing between the two
will happen. In our case, since the remaining terms cannot be very
large due to the small $\xi$, the amplitude of $S_c^-$ will be
suppressed, and in a considerable region of large scales, we can
also obtain a scale-invariant spectrum \footnote{Actually, there may
exist some fine-tuning between the amplitude of $S_c^-$ and
$k^2S_c^-$, and if the former is too large, we might not get a
scale-invariant power spectrum. Although one may think from naive
intuition that it will not be that severe because of the small
$\xi$, a careful comparison should be needed in order to have a safe
theory. We would investigate this point in detail in a future
work.}. Nevertheless, for extremely large scales, the contribution
of $S_c^-$ of zeroth order in $k$ will become non-negligible, and
the spectrum will have a red-tilt.
\subsubsection{numerical results}
\begin{figure}[htbp]
\includegraphics[scale=0.3]{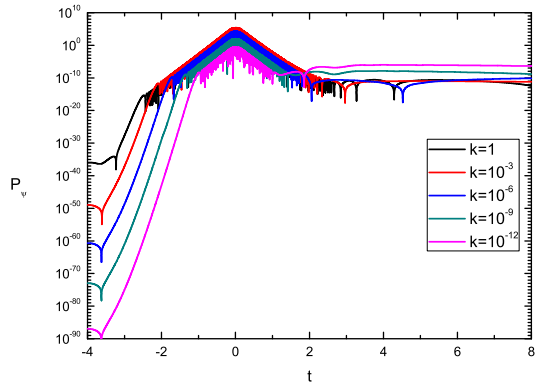}
\caption{(Color online) Results of the numerical evolutions of the
curvature perturbations with different comoving wavenumbers $k$ in
non-minimal coupling matter bounce as function of cosmic time $t$.
The initial values of the background parameters are the same as in
Figure 1. The units of the time axis are $M_{rec}^{-1}$, the
comoving wavenumber $k$ is unity for $k=m_{pl}$. }\label{spectrum}
\end{figure}

In order to support our analysis above, we also perform the
numerical calculation for the perturbations. Fig. \ref{spectrum} is
the numerical results for the dependence of the metric power
spectrum ${\cal P}_\Psi$ on the cosmic time $t$ for different
comoving $k$ modes, where we use the normal definition of power
spectrum as:\be {\cal
P}_\Psi(k)\equiv\frac{k^3}{2\pi^2}|\Psi|^2\simeq\frac{k^3}{2\pi^2}|D_{e}^{+}|^2~.\ee
We choose the zero point on the horizontal axis to be the bouncing
point, and set the initial conditions to be Bunch-Davies vacuum. One
can see from our plot that before the bounce point, the perturbation
is dominated by their growing modes, and after the bounce, it is
dominated by the constant modes, which fits the analytical results
very well. As for the $k$ dependence, we see that for medium $k$
modes, the power spectrum takes on scale invariance while in the
extremely small $k$ modes where $k\lesssim10^{-6}m_{pl}$, the
spectrum will present a slightly red tilt. This is because the non
trivial inheritance of the growing mode in contracting phase to the
constant mode in expanding phase at zeroth order of $k$ due to the
non-minimal coupling effects. Since the scale variance happens only
in extreme large scales, we expect that it could be tested in the
future observational data.

\begin{figure}[htbp]
\includegraphics[scale=0.3]{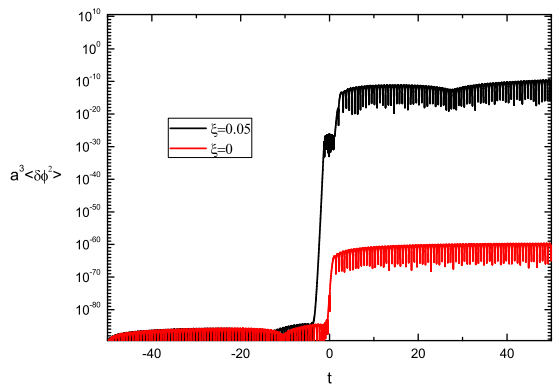}
\caption{(Color online) Results of the numerical evolutions of the
field particle production with and without non-minimal coupling term
for matter bounce. The initial values of the background parameters
are the same as in Figure 1.}\label{reheating}
\end{figure}

Furthermore, we calculate numerically the evolution of the particle
production $\delta\phi$ with respect to $t$ and plot the result in
Fig. \ref{reheating}. Particle production is expected in the region
where the squared value of momentum scales of the perturbation modes
are larger than the potential so that the WKB approximation becomes
valid, such as the reheating process at the end of inflation
\cite{Albrecht:1982mp}. Here at the bouncing point crossing, there
is also possibilities to produce particles. In the usual case, we
can define another variable $\varphi\equiv a^\frac{3}{2}\delta\phi$,
which satisfies equation of motion as: \be
\ddot\varphi+\omega_k^2\varphi=...,\ee where
$\omega_k^2=\frac{k^2}{a^2(t)}+m^2-\frac{3}{2}\dot
H-\frac{9}{4}H^2$. From above we can see that, although the first
two terms are positive-determined, there will remain some additional
terms due to the geometry of the universe, which would cause
$\omega_k^2$ to be less than zero. If this is the case, the field
will have tachyonic behavior and will blow up with particle
production (which can be called as tachyonic reheating
\cite{Greene:1998pb}). However, if there is a non-minimal coupling
term with a negative coupling constant $\xi$, this effect will get
larger since $\omega_k^2$ gets a more negative value. This can be
seen from the modified equation of motion which becomes: \be
\ddot\varphi+[\frac{k^2}{a^2(t)}+m^2-\frac{3}{2}\dot
H-\frac{9}{4}H^2-\xi R]\varphi=... \ee This is usually called
geometric reheating \cite{Bassett:1997az}. In Fig. \ref{reheating}
we can see that indeed the particle production gets much enhancement
in the non-minimal coupling case than its minimal coupling
counterparts. This is another interesting result in our case which
are expected to be tested by the future experiments.

\section{conclusions and discussions}
Non-minimal coupling is a very popular subject in cosmology and has
been widely studied in the literature. In this paper, we have
investigated the possibility of generating a matter bounce by the
Lee-Wick type scalar field with a non-minimal coupling term
involved, and studied its perturbations. As in the previous work
done in \cite{Cai:2008qw}, a bounce was obtained when the
non-minimal coupling is not too large to have unexpected effects.
However, as it gives correction to the effective mass of the field,
a short period of deflation/inflation before/after the bounce will
happen.

Using the standard techniques of calculating perturbations of bounce
developed from previous works, we have calculated the perturbations
of the model in detail. One of the big differences from previous one
for minimal coupling is that the non-minimal coupling to gravity
causes the difference between the two scalar perturbations $\Phi$
and $\Psi$, which will cause very interesting consequences. We have
obtained the solution for each stage of evolution and compiled them
together using proper matching conditions. We have found that the
final dominant mode can be inherited nontrivially from the
subdominant mode in the zero-th order of wave number $k$ due to this
difference. This will lead to the red-tilt of the power spectrum at
very large scales, which can be justified in future observation
\footnote{Another interesting case of tilting the spectrum would be
to consider closed universe and have slow bounce of which $\dot
H\rightarrow0$ at the bounce point while higher time derivatives of
$H$ drive the bounce, in which case the perturbations will transfer
through the bounce via a scale-dependent way. Such works were
discussed in e.g. \cite{Martin:2003bp}. We thank the anonymous
referee for point this to us.}.

Another result we obtained is that due to the non-minimal coupling,
the particle production of the scalar will get greatly enhanced. In
the region of validity of the WKB approximation when the quantum
effects become significant, the field will get excited and produce
particles. This is very common in reheating process. When there is a
negative effective mass squared term in the field equation, the
particle production will get enhanced due to tachyonic resonance. In
usual cases, the averaged value of perturbations of the scalar field
will get an enhancement through bounce time due to geometrical
effects. When a non-minimal coupling term is involved in the field
equation, the mass of the field will be corrected. In the case of
negative coupling coefficient, the field will get a more negative
effective mass squared, and the particle production will get more
efficiently enlarged. This is also very interesting phenomenon and
can be tested by experiments on particle physics.

Finally we would like to close with some remarks on future works.
The bounce with non-minimal coupling is very interesting topic since
in the early epoch, Einstein's gravity is likely to be modified,
which can give some valuable effects to the evolution of our
universe. It is deserved to pay attention to this project both from
the theoretical side such as the perturbation, non-Gaussianity,
effects on CMB and so on, and from observational side, such as the
constraints, confirmation or even exclusion from experimental data.
All these works are expected to take on in the future.
\section*{Acknowledgments}

One of us (T.Q.) thanks Prof. Robert Brandenberger, Prof. Yunsong
Piao, Prof. Shinji Tsujikawa, Yifu Cai and Jun Zhang for useful
suggestions at the beginning of the work. This research is supported
in parts by the National Science Council of R.O.C. under Grant No.
NSC96-2112-M-033-004-MY3 and No. NSC97-2811-033-003 and by the
National Center for Theoretical Science.

\appendix
\section{Hubble parameter in Jordan and Einstein frames}

This appendix is set to claim the relation between Hubble parameter
in the two frames used in the text. As is well known, the
differentiations of comoving time $d\eta$ are the same in the two
frames, while that of cosmic time $dt$ are not. By definition, we
have $d\hat{t}=\sqrt{F}dt$ where $\hat{t}$ denotes the cosmic time
in Einstein frame. Due to this difference, one should be careful
when using variables with derivatives in terms of $t$ such as Hubble
parameter $H$. As is already mentioned in the previous sections, the
Hubble parameter defined in Jordan frame is: \be H\equiv\frac{\dot
a}{a}=\frac{da}{adt}~.\ee However, we can also define a Hubble
parameter in Einstein frame, which is: \be \hat{H}\equiv
\frac{\stackrel{\circ}{\hat{a}}}{\hat{a}}=\frac{d\hat{a}}{\hat{a}d\hat{t}}~,\ee
where we have introduced the symbol $``\stackrel{\circ}{}''$ to
represent the cosmic time derivative in Einstein frame:
$\stackrel{\circ}{}=\frac{d}{d\hat{t}}$,
$\stackrel{\circ\circ}{}=\frac{d^2}{d\hat{t}^2}$, etc. We can see
from the definitions that the two Hubble parameters have the same
type of relation to the comoving Hubble parameters in their frames:
${\cal H}=aH$ and $\hat{\cal H}=\hat{a}\hat{H}$. Reminding the
relation $\hat{a}=a\sqrt{F}$, it is easy to find the description of
$\hat{H}$ in terms of $H$ and $F$: \bea
\hat{H}&=&\frac{d(a\sqrt{F})}{aFdt}~\nonumber\\
&=&\frac{H}{\sqrt{F}}-(\frac{1}{\sqrt{F}})\dot{}~.\eea Furthermore,
we can derive its derivatives with respect to cosmic time $\hat{t}$
which are: \bea \stackrel{\circ}{\hat{H}}&=&\frac{\dot
H}{F}+H(\frac{1}{2F})\dot{}-\frac{1}{\sqrt{F}}(\frac{1}{\sqrt{F}})\ddot{}~,\\
\stackrel{\circ\circ}{\hat{H}}&=&\frac{\ddot H}{F\sqrt{F}}+\dot
H(\frac{1}{F\sqrt{F}})\dot{}+\frac{H}{\sqrt{F}}(\frac{1}{2F})\ddot{}-\frac{1}{F}(\frac{1}{\sqrt{F}})^\dddot{}-(\frac{1}{2F})\dot{}(\frac{1}{\sqrt{F}})\ddot{}~.\eea

\section{coefficients for the perturbation modes}
This appendix indicates all the coefficients that appear in
(\ref{final}). These are calculated using the Hwang-Vishniac
(Dueruelle-Mukhanov) matching conditions
\cite{Hwang:1991an,Deruelle:1995kd}: \be
^{(0)}G_{c}^{+}=\frac{\frac{2}{3aFt^{\frac{5}{3}}A_{c}^{+}}+B_{c}^{+}F[(\frac{20\epsilon}{9aFt^{\frac{8}{3}}}+\frac{2\dot{A_{c}^{+}}}{aFt^{\frac{5}{3}}A_{c}^{+}})H+(16\sqrt{\frac{1}{3}}\frac{\epsilon\kappa\xi\phi}{aF^{2}t^{\frac{5}{3}}}+\frac{2}{aFt^{\frac{5}{3}}}-\frac{4\epsilon}{3aFt^{\frac{5}{3}}}-\frac{4C_{c}^{+}}{aF^{2}t^{\frac{5}{3}}A_{c}^{+}})H^{2}]}{(-3B_{c}^{+}\dot{A_{c}^{+}}FH+3A_{c}^{+}\dot{B_{c}^{+}}FH+6B_{c}^{+}C_{c}^{+}H^{2}-6A_{c}^{+}E_{c}^{+}H^{2})}\Big|_{t_c}~,\ee
\be
^{(0)}K_{c}^{+}=\frac{\frac{6}{5\kappa^{2}A_{c}^{+}}-B_{c}^{+}F[\frac{18\dot{A_{c}^{+}}H}{5\kappa^{2}A_{c}^{+}}+(2.4\frac{\epsilon}{\kappa^{2}}+6.4\sqrt{3}\frac{\epsilon\xi\phi}{F\kappa}+\frac{18}{5\kappa^{2}}-\frac{36C_{c}^{+}}{5\kappa^{2}A_{c}^{+}F})H^{2}]}{(-3B_{c}^{+}\dot{A_{c}^{+}}FH+3A_{c}^{+}\dot{B_{c}^{+}}FH+6B_{c}^{+}C_{c}^{+}H^{2}-6A_{c}^{+}E_{c}^{+}H^{2})}\Big|_{t_c}~,\ee
\be
^{(1)}G_{c}^{+}=\frac{B_{c}^{+}F(\frac{4\epsilon}{9aFt^{\frac{5}{3}}}-\frac{2}{3aFt^{\frac{5}{3}}})}{(-3B_{c}^{+}\dot{A_{c}^{+}}FH+3A_{c}^{+}\dot{B_{c}^{+}}FH+6B_{c}^{+}C_{c}^{+}H^{2}-6A_{c}^{+}E_{c}^{+}H^{2})}\Big|_{t_c}~,\ee
\be
^{(1)}K_{c}^{+}=\frac{B_{c}^{+}F(\frac{4\epsilon}{5\kappa^{2}}-\frac{6}{5\kappa^{2}})}{(-3B_{c}^{+}\dot{A_{c}^{+}}FH+3A_{c}^{+}\dot{B_{c}^{+}}FH+6B_{c}^{+}C_{c}^{+}H^{2}-6A_{c}^{+}E_{c}^{+}H^{2})}\Big|_{t_c}~,\ee
\be
^{(0)}M_{c}^{+}=\frac{-A_{c}^{+}F[(\frac{20\epsilon}{9aFt^{\frac{8}{3}}}+\frac{2\dot{A_{c}^{+}}}{aFt^{\frac{5}{3}}A_{c}^{+}})H+(16\sqrt{\frac{1}{3}}\frac{\epsilon\kappa\xi\phi}{aF^{2}t^{\frac{5}{3}}}+\frac{2}{aFt^{\frac{5}{3}}}-\frac{4\epsilon}{3aFt^{\frac{5}{3}}}-\frac{4C_{c}^{+}}{aFt^{\frac{5}{3}}})H^{2}]}{(-3B_{c}^{+}\dot{A_{c}^{+}}FH+3A_{c}^{+}\dot{B_{c}^{+}}FH+6B_{c}^{+}C_{c}^{+}H^{2}-6A_{c}^{+}E_{c}^{+}H^{2})}\Big|_{t_c}~,\ee
\be
^{(0)}N_{c}^{+}=\frac{A_{c}^{+}F[\frac{18\dot{A_{c}^{+}}H}{5\kappa^{2}A_{c}^{+}}-(2.4\frac{\epsilon}{\kappa^{2}}+6.4\sqrt{3}\frac{\epsilon\xi\phi}{F\kappa}-\frac{18}{5\kappa^{2}}+\frac{36C_{c}^{+}}{5\kappa^{2}A_{c}^{+}F})H^{2}]}{(-3B_{c}^{+}\dot{A_{c}^{+}}FH+3A_{c}^{+}\dot{B_{c}^{+}}FH+6B_{c}^{+}C_{c}^{+}H^{2}-6A_{c}^{+}E_{c}^{+}H^{2})}\Big|_{t_c}~,\ee
\be
^{(1)}M_{c}^{+}=\frac{A_{c}^{+}F(\frac{4\epsilon}{9aFt^{\frac{5}{3}}}-\frac{2}{3aFt^{\frac{5}{3}}})}{(-3B_{c}^{+}\dot{A_{c}^{+}}FH+3A_{c}^{+}\dot{B_{c}^{+}}FH+6B_{c}^{+}C_{c}^{+}H^{2}-6A_{c}^{+}E_{c}^{+}H^{2})}\Big|_{t_c}~,\ee
\be
^{(1)}N_{c}^{+}=\frac{A_{c}^{+}F(\frac{4\epsilon}{5\kappa^{2}}-\frac{6}{5\kappa^{2}})}{(-3B_{c}^{+}\dot{A_{c}^{+}}FH+3A_{c}^{+}\dot{B_{c}^{+}}FH+6B_{c}^{+}C_{c}^{+}H^{2}-6A_{c}^{+}E_{c}^{+}H^{2})}\Big|_{t_c}~,\ee
\be
^{(0)}M_{e}^{+}=\frac{\kappa^{2}(2\dot{A_{e}^{-}}Ft^{\frac{8}{3}}-4C_{e}^{-}Ht^{\frac{8}{3}}+\frac{20}{9}t^{\frac{5}{3}}\epsilon
A_{e}^{-}F+2A_{e}^{-}HFt^{\frac{8}{3}}-\frac{4}{3}\epsilon
A_{e}^{-}HFt^{\frac{8}{3}}+8A_{e}^{-}\kappa^{2}\xi\phi
Ht^{\frac{8}{3}}\epsilon\sqrt{\frac{4}{3\kappa^{2}}})}{\epsilon(\frac{8}{3}Ft^{\frac{5}{3}}+16\kappa^{2}\xi\phi
Ht^{\frac{8}{3}}\sqrt{\frac{4}{3\kappa^{2}}})}\Big|_{t_e}~,\ee \be
^{(0)}N_{e}^{+}=\frac{\kappa^{2}(2\dot{B_{e}^{-}}Ft^{\frac{8}{3}}-4E_{e}^{-}Ht^{\frac{8}{3}}+\frac{20}{9}t^{\frac{5}{3}}\epsilon
B_{e}^{-}F+2B_{e}^{-}HFt^{\frac{8}{3}}-\frac{4}{3}\epsilon
B_{e}^{-}HFt^{\frac{8}{3}}+8B_{e}^{-}\kappa^{2}\xi\phi
Ht^{\frac{8}{3}}\epsilon\sqrt{\frac{4}{3\kappa^{2}}})}{\epsilon(\frac{8}{3}Ft^{\frac{5}{3}}+16\kappa^{2}\xi\phi
Ht^{\frac{8}{3}}\sqrt{\frac{4}{3\kappa^{2}}})}\Big|_{t_e}~,\ee \be
^{(1)}M_{e}^{+}=\frac{\kappa^{2}Ft{}^{\frac{8}{3}}(-\frac{2}{3}+\frac{4}{9}\epsilon)A_{e}^{-}}{H\epsilon(\frac{8}{3}Ft^{\frac{5}{3}}+16\kappa^{2}\xi\phi
Ht^{\frac{8}{3}}\sqrt{\frac{4}{3\kappa^{2}}})}\Big|_{t_e}~,\ee \be
^{(1)}N_{e}^{+}=\frac{\kappa^{2}Ft{}^{\frac{8}{3}}(-\frac{2}{3}+\frac{4}{9}\epsilon)B_{e}^{-}}{H\epsilon(\frac{8}{3}Ft^{\frac{5}{3}}+16\kappa^{2}\xi\phi
Ht^{\frac{8}{3}}\sqrt{\frac{4}{3\kappa^{2}}})}\Big|_{t_e}~,\ee where
\be A_c^+=\frac{1}{a_1\epsilon
(t-t_{\ast})^{1+\frac{1}{\epsilon}}[1-\frac{4\xi\phi_0}{3}(\ln{(t-t_{\ast})})^2]}~,~B_c^+=\frac{2\epsilon}{\kappa^2}\Bigl(\frac{1}{1+\epsilon}+\frac{8\xi\phi_0[\epsilon-(1+\epsilon)\ln{(t-t_{\ast})}]}{3(1+\epsilon)^3[1-\frac{4\xi\phi_0}{3}(\ln{(t-t_{\ast})})^2]}\Bigr)~,\ee
\be
C_c^+=\frac{\frac{4}{\sqrt{3\kappa^2}}\kappa^2\xi\phi}{a_{1}(t-t_{\ast})^{1+\frac{1}{\epsilon}}[1-\frac{4\xi\phi_{0}}{3}(\ln{(t-t_{\ast})})^{2}]}~,\ee
\be
E_c^+=-\frac{8\epsilon\xi\phi\{(1+\epsilon)^{2}[1-\frac{4\xi\phi_{0}}{3}(\ln{(t-t_{\ast})})^{2}]-\frac{8\epsilon\xi\phi_{0}}{3}[\epsilon+(1+\epsilon)\ln{(t-t_{\ast})}]\}}{\sqrt{3\kappa^{2}}(1+\epsilon)^{3}[1-\frac{4\xi\phi_{0}}{3}(\ln{(t-t_{\ast})})^{2}]}~,\ee
\be A_e^-=\frac{1}{a_2\epsilon
(t-t_{\dag})^{1+\frac{1}{\epsilon}}[1-\frac{4\xi\phi_0}{3}(\ln{(t-t_{\dag})})^2]}~,~B_e^-=\frac{2\epsilon}{\kappa^2}\Bigl(\frac{1}{1+\epsilon}+\frac{8\xi\phi_0[\epsilon-(1+\epsilon)\ln{(t-t_{\dag})}]}{3(1+\epsilon)^3[1-\frac{4\xi\phi_0}{3}(\ln{(t-t_{\dag})})^2]}\Bigr)~,\ee
\be
C_e^-=\frac{\frac{4}{\sqrt{3\kappa^2}}\kappa^2\xi\phi}{a_{2}(t-t_{\dag})^{1+\frac{1}{\epsilon}}[1-\frac{4\xi\phi_{0}}{3}(\ln{(t-t_{\dag})})^{2}]}~,\ee
\be
E_e^-=-\frac{8\epsilon\xi\phi\{(1+\epsilon)^{2}[1-\frac{4\xi\phi_{0}}{3}(\ln{(t-t_{\dag})})^{2}]-\frac{8\epsilon\xi\phi_{0}}{3}[\epsilon+(1+\epsilon)\ln{(t-t_{\dag})}]\}}{\sqrt{3\kappa^{2}}(1+\epsilon)^{3}[1-\frac{4\xi\phi_{0}}{3}(\ln{(t-t_{\dag})})^{2}]}~.\ee
The constants $t_\ast$ and $t_\dag$ are set to guarantee the
continuity of $H$ at the joint point $t_c$ and $t_e$, and one can
straightforwardly check that in the limit of $\xi=0$, we have
$^{(0)}M_{c}^{+}$ and $^{(0)}M_{e}^{+}$=0, and a scale-invariant
power spectrum can be obtained for the whole region of large scales.

\end{document}